%% file: atlas_dndetaHI_plb.tex
\RequirePackage{lineno}
\documentclass[preprint]{elsarticle}
\pdfoutput=1
\DeclareGraphicsExtensions{.pdf}

\usepackage{amssymb}
\usepackage{lineno}
\usepackage{slashed}
\usepackage{graphicx}
\newcommand{\sqrts}{\mbox{$\sqrt{s_{_\mathrm{NN}}}$}}
\usepackage{atlasphysics}
\usepackage{hyperref}

\newcommand{\Nch}{\mbox{$N_{\mathrm{ch}}$}}
\newcommand{\Nevt}{\mbox{$N_{\mathrm{evt}}$}}
\newcommand{\Ntptr}{\mbox{$N_{\mathrm{2p}}$}}
\newcommand{\Ntptrt}{N_{\mathrm{2p}}}
\newcommand{\Npart}{\mbox{$\langle N_{\mathrm{part}} \rangle$}}

\newcommand{\Nraw}{\mbox{$N^{\mathrm{raw}}$}}

\newcommand{\nucnuc}{\mbox{A+A}}
\newcommand{\AuAu}{\mbox{Au+Au}}
\newcommand{\PbPb}{\mbox{Pb+Pb}}
\newcommand{\pp}{\mbox{p+p}}
\newcommand{\kt}{\mbox{$k_{\mathrm{T}}$}}
\newcommand{\occ}{\mbox{$\mathcal{O}$}}
\newcommand{\Rmc}{\mbox{$\Delta {\mathcal R}$}}

\newcommand{\dndeta}{\mbox{$d\Nch/d\eta$}}
\newcommand{\dndetaperpair}{\mbox{$d\Nch/d\eta/(\Npart/2)$}}
 
\newcommand{\dndetamid}{\mbox{$d\Nch/d\eta|_{\eta=0}$}}
\newcommand{\sumet}{\mbox{$\Sigma E_{\mathrm{T}}$}}

\journal{Physics Letters B}

\begin{document}

\title{
\vspace{-0.5cm} \flushright{\normalsize{CERN-PH-EP-2011-118}} \\ 
\center{Measurement of the centrality dependence of the charged
  particle pseudorapidity distribution in lead-lead collisions 
  at $\sqrts = 2.76$~TeV with the ATLAS detector}
}
\author{The ATLAS Collaboration}

\begin{abstract}
The ATLAS experiment at the LHC has measured the centrality dependence
of charged particle pseudorapidity distributions over $|\eta| < 2$ in
\mbox{lead-lead} collisions at a nucleon-nucleon centre-of-mass energy
of $\sqrts=2.76$~TeV.
In order to include particles with transverse momentum as low as 30 MeV, the
data were recorded with the central solenoid magnet off. Charged particles were reconstructed
with two algorithms (2-point ``tracklets'' and full tracks) 
using information from the pixel detector only.
The lead-lead collision centrality was
characterized by the total transverse energy in the forward
calorimeter in the range $3.2 < |\eta| < 4.9$. Measurements are presented of
the per-event charged particle pseudorapidity distribution, \dndeta, 
and the average charged particle multiplicity in the pseudorapidity 
interval $|\eta|<0.5$ in several intervals of collision
centrality. The results are compared to
previous mid-rapidity measurements at the LHC and RHIC. The variation of the
mid-rapidity charged particle yield per colliding nucleon pair with
the number of participants is consistent with lower
\sqrts\ results. The shape of the \dndeta\ distribution is found to be
independent of centrality within the systematic uncertainties 
of the measurement.

\end{abstract}
\maketitle 


\section{Introduction}
Collisions of lead (Pb) ions at the Large Hadron Collider provide an
opportunity to study strongly interacting matter at the highest
temperatures ever created in the laboratory~\cite{LHCHI}. Measurements
of the centrality dependence of charged particle multiplicities and of
charged particle pseudorapidity densities in such ultra-relativistic
nucleus-nucleus (\nucnuc) collisions provide essential information on
the initial particle or entropy production and subsequent evolution in
the created hot, dense matter. Results from the Relativistic Heavy Ion
Collider (RHIC) over the centre-of-mass energy range from 19.6 to 200
GeV indicate that the multiplicity of charged particles per colliding
nucleon pair has a mild dependence on the collision centrality and
that the pseudorapidity dependence of the charged particle yield near
mid-rapidity is essentially centrality independent~\cite{Alver:2010ck}. 
The weak variation of the multiplicity per colliding nucleon pair with
centrality at RHIC was initially found to be inconsistent with models
such as HIJING~\cite{Wang:1991hta} which includes a mixture of soft and hard
scattering processes with a $p_T$ cutoff on the hard scattering
contribution at 2~GeV, or with a beam-energy-dependent cutoff in a 
more recent version~\cite{Deng:2010mv}. In contrast, calculations based on parton
saturation invoking \kt\ factorization were able to reproduce both
the shape and centrality dependence of the RHIC charged particle
pseudorapidity distributions~\cite{ALbacete:2010ad, Levin:2010dw}. 
However, more recent theoretical studies
indicate that \kt\ factorization may not be applicable to
nucleus-nucleus collisions, and improved soft$+$hard models may be
able to describe RHIC multiplicity measurements. 
At the same time, older hydrodynamical models (e.g. Ref~\cite{Carruthers:1973ws})
have had some success describing the energy dependence of the total
multiplicity as well as rapidity distributions of identified hadrons,
although their domain of applicability is still not fully established.

Detailed measurements of the centrality dependence of charged particle
multiplicities and pseudorapidity distributions at the LHC together
with the earlier RHIC measurements could provide essential insight on
the physics responsible for bulk particle production in
ultra-relativistic nuclear collisions. Because hard scattering rates
increase rapidly with centrality and $\sqrts$, the combined RHIC and
LHC measurements should provide a strong constraint on the
contribution of hard scattering processes to inclusive hadron
production subject to uncertainties regarding the shadowing of nuclear
parton distributions at low $x$. Measurements at the LHC can also
provide a valuable test of recent parton saturation
calculations that still claim to be able to describe inclusive
particle production in ultra-relativistic nuclear collisions~\cite{ALbacete:2010ad, Levin:2010dw}.
Previous measurements at the LHC
\cite{Collaboration:2010cz,CMS:2011mult} have already started
addressing some of the physics raised above. In particular, those
earlier measurements found a rapid rise in the particle multiplicity
at the LHC compared to naive extrapolations of RHIC measurements and a
variation of mid-rapidity charged particle multiplicity with
centrality similar to that observed at RHIC.

This paper
presents the results of ATLAS~\cite{AtlasDetector} measurements of
the per-event charged particle
pseudorapidity distribution, \dndeta, 
in $\sqrts = 2.76$~TeV \PbPb\ collisions
over $|\eta| < 2$ and as a function of collision centrality with the
goal of testing and extending
the results of the previous LHC measurements.
In this paper, $\Nch$ denotes the per-event number of charged
primary particles measured in an interval of $\eta$, 
which is the particle pseudorapidity~\footnote{ATLAS uses a right-handed coordinate system with its origin at the
nominal interaction point (IP) in the centre of the detector and the
$z$-axis along the beam pipe. The $x$-axis points from the IP to the
centre of the LHC ring, and the $y$ axis points upward. Cylindrical
coordinates $(r,\phi)$ are used in the transverse plane, $\phi$ being
the azimuthal angle around the beam pipe. The pseudorapidity is
defined in terms of the polar angle $\theta$ as
$\eta=-\ln\tan(\theta/2)$.}.
The measurement was performed with the solenoid off, thereby allowing 
detection of charged particles down to very low transverse momenta
($\pt \sim 30$~MeV). 

\section{Experimental setup and event selection}
The measurements presented here were obtained using the ATLAS inner
detector \cite{atlasdetector} which
contains both silicon pixel and silicon strip detectors and the
ATLAS forward calorimeters. The charged particle multiplicity is measured 
using the pixel detector~\cite{pixel} which consists of
three layers of pixel staves in the barrel region, inclined at an
angle of $20^\circ$, at radii of 50.5, 88.5, and 122.5~mm from the
nominal beam axis. The typical pixel size is $50~{\rm \mu m} \times
400~{\rm \mu m} $ in $\phi-z$, and an average occupancy of about $
0.5$\% is observed for the innermost pixel layer in central
\PbPb\ collisions. To limit low-\pt\ multiple scattering losses in
detector material, the measurement has been restricted to the barrel
portion of the pixel detector, corresponding to pseudorapidity values
in the range $|\eta|<2$. Collision vertex positions were obtained by
full reconstruction of nominally straight charged particle
trajectories in the pixel and silicon strip detectors followed by
reconstruction of a single collision vertex from the full set of
particle trajectories. To maintain uniform acceptance of the pixel
detector for the multiplicity measurement the vertex was required to
lie within 50~mm of the nominal centre of the ATLAS detector in the
longitudinal direction. 

The data for the measurements presented here were collected with a
minimum-bias trigger.  This required a coincidence in either the two
minimum-bias trigger scintillator (MBTS) detectors, located at $\pm
3.56~$m from the interaction centre and covering $2.1 < |\eta| < 3.9$,
or two zero-degree calorimeters (ZDCs), located at $\pm 140~$m from
the interaction centre and covering $|\eta|>8.3$. The threshold on the
analog energy sum in each ZDC was set below the single neutron
peak. The offline analysis required the time difference between the
two MBTS detectors to be $|\Delta t|<3$~ns to eliminate upstream
beam-gas interactions, a ZDC coincidence to efficiently reject
photo-nuclear events~\cite{Djuvsland:2010qs}, and a reconstructed
vertex satisfying the selection described above.  The measurements
presented in this paper were obtained from a 10~hour data-taking run
corresponding to an integrated luminosity of approximately $480~{\rm
  mb^{-1}}$. A total of 1631525 events passed the trigger, vertex, and
offline selections.

\section{Centrality}

In heavy ion collisions, ``centrality'' reflects the overlap  
volume of the two colliding nuclei, controlled by the classical impact
parameter. That overlap volume is closely related to the number
of ``participants'', the nucleons which scatter inelastically in 
each nuclear collision.  While the number of participants,
$N_{\mathrm{part}}$, cannot be measured for a single collision,
previous studies at RHIC and the SPS have demonstrated that the
multiplicity and transverse energy of the produced particles are
strongly correlated with $N_{\mathrm{part}}$.  Because of this, the
average number of participants can be accurately estimated from a
selected fraction of the multiplicity or transverse energy
distribution~\cite{Miller:2007ri}. 
In ATLAS, the \PbPb\ collision centrality is measured using
the summed transverse energy (\sumet) in the
forward calorimeter (FCal) over the pseudorapidity range $3.2 <
|\eta| < 4.9$, calibrated at the electromagnetic energy scale.
An analysis of the FCal \sumet\ distribution after application
of all trigger and 
selection requirements gives an estimate of the fraction
of the sampled non-Coulomb inelastic cross section  
of $f = 98 \pm 2\%$. This estimate was derived from
comparisons of the measured FCal \sumet\ distribution with 
a simulated \sumet\ distribution. 
The simulated distribution was obtained from 
a convolution of $\sqrt{s}=2.76$ TeV proton-proton data with a Monte Carlo (MC) 
Glauber calculation~\cite{Miller:2007ri,Alver:2008aq} of the number of effective
nucleon-nucleon collisions.  This quantity 
was calculated as a linear combination of
the number of participants and the number of binary collisions, similar to
what was done in a previous analysis~\cite{:2010px}.
The value of $f$ and its uncertainty was estimated by
systematically varying the effect
of trigger and event selection inefficiencies as well as backgrounds 
in the most peripheral $\sumet$ interval.
 This was done by artificially injecting and removing counts in that interval in order to achieve the best agreement between the measured and simulated distributions.
The estimate of $f$ was made after removal of a 1\% background contamination 
in the most peripheral events that was evaluated using comparisons of solenoid magnet-on 
and solenoid magnet-off data and which was attributed to photo-nuclear events.

For the results presented in this paper, the minimum-bias FCal \sumet\ distribution
was divided into centrality intervals according to the following percentiles: 
10\% intervals over 0-80\%,
5\% intervals over $20-80\%$ and 2\% intervals over $0-20\%$.
By convention, the 0-10\% centrality interval refers to the 10\%
most central events -- the events with the highest
\sumet\ values\ -- and increasing percentiles refer to 
events with successively lower $\sumet$. 
The average number of participants, $\Npart$, was evaluated
for each of the experimental centrality intervals by dividing the Glauber model
\sumet\ distribution into the same percentile centrality intervals used 
for the data and evaluating the average number of participants of the 
Glauber MC events contributing to a given interval.
This procedure incorporates more realistic fluctuations into the estimation of $\Npart$ than
would be achieved by binning in either $N_{\mathrm{part}}$ itself or in the classical impact parameter. 
The systematic errors on $\Npart$ 
were evaluated from the quoted uncertainty on
$f$ and the known uncertainties in the nuclear density parameters as well
as the assumed total inelastic nucleon-nucleon cross section of
$\sigma_{\mathrm{NN}} = 64\pm 5$~mb~\cite{Nakamura:2010zzi}.

\section{Reconstruction of charged particle multiplicity}
In the offline analysis, adjacent hits in the pixel
  modules were grouped into clusters using standard techniques.
Two methods were, then,
used to reconstruct charged particles from the pixel clusters.
In one method, a Kalman Filter-based tracking algorithm, 
similar to that deployed in proton-proton collisions~\cite{Newtracking1}, 
was applied only to the pixel layers (``pixel tracks''). 
The other method, the ``two-point tracklet'' algorithm, 
used the reconstructed primary vertex and clusters on the first pixel
layer to define a search region for clusters in the second layer
consistent with a nominally straight track. Candidate tracklets were
required to have deviations between projected and measured cluster
positions in the second pixel layer in pseudorapidity and azimuth, $\Delta \eta$ and $\Delta
\phi$, respectively, satisfying
\begin{equation}
\Delta {\mathcal R} \equiv \frac{1}{\sqrt{2}} \sqrt{\left(\frac{\Delta \eta}{\sigma_\eta(\eta)}\right)^2 + 
\left(\frac{\Delta \phi}{\sigma_\phi(\eta)}\right)^2} <  3.
\label{eq:trackletcuts}
\end{equation}
The widths 
of the $\Delta \eta$ and $\Delta \phi$ distributions
characterized by the pseudorapidity-dependent resolutions $\sigma_\eta(\eta)$ and
$\sigma_\phi(\eta)$ were obtained from the MC simulations described below.
The $\eta$ and $\phi$ values of the reconstructed tracklets 
were determined using the
cluster position on the first layer and the primary vertex position.
The two-point tracklet analysis excluded
clusters with low energy deposits inconsistent with minimum-ionizing particles
originating at the primary vertex.  It also excluded duplicate clusters resulting
from the overlap of the pixel modules in $\phi$ and 
from a small set of pixels at the centres of the pixel modules
that share readout channels~\cite{pixel}.

The high charged particle multiplicity in \PbPb\ collisions can 
generate misidentified tracks and/or two-point tracklets when only two or three
measurements are made on each trajectory. The misidentified 
contributions have been evaluated using the MC
  studies described below, but to check the
  MC results, an independent,
data-driven estimate of misidentified two-point tracklets was obtained using a variant of the
  two-point tracklet algorithm. In the default
two-point tracklet analysis, referred to as ``Method 1'', at most
one tracklet was reconstructed for a given cluster on the first pixel
layer. If multiple clusters on the second pixel layer
fell within the search region defined in
Equation~\ref{eq:trackletcuts}, the closest cluster to the projected
position was chosen. This method limits, but does not eliminate, the generation of
misidentified tracklets.  A second implementation of the two-point tracklet
algorithm, referred to as ``Method 2'', produced tracklets for 
all combinations of clusters on the two layers consistent with 
the search region. Using Method~2, the rate of false tracklets
resulting from random combinations of clusters was
estimated by performing the same analysis but with the clusters
on the second layer having their $z$ positions inverted around the
primary vertex and their azimuthal
angles inverted, $\phi \rightarrow \pi-\phi$. The tracklet yield from
this ``flipped'' analysis was then subtracted from the
proper tracklet yield event-by-event to obtain the estimated yield of
true tracklets, 
\begin{equation}
\Ntptrt(\eta) = \Ntptrt^{\mathrm{ev}}(\eta) - \Ntptrt^{\mathrm{fl}} (\eta),
\label{eq:flipsub}
\end{equation}
where $ \Ntptrt^{\mathrm{ev}}$ represents the yield of two-point tracklets using
Method 2 
and $\Ntptrt^{\mathrm{fl}}$
represents the yield obtained by flipping the clusters in the second
pixel layer. For the 0-10$\%$ centrality interval, the flipped yield is
about 50$\%$ of the unflipped yield in the $|\eta|<0.5$ region.

The response of the detector to the charged
particles produced in \PbPb\ collisions and the performance of the
track and tracklet methods was evaluated by MC 
simulations of \PbPb\ collisions using the
HIJING~\cite{Wang:1991hta} event generator followed by
GEANT4~\cite{Agostinelli:2002hh} simulations of the detector response~\cite{ATLASSim}.
The resulting events were then reconstructed and analyzed using
the full offline analysis 
chain that was applied to the experimental data. 
HIJING events were generated without jet quenching and with an
unbiased impact parameter distribution. Impact parameter and
\pt-dependent elliptic flow was imposed on the HIJING events after
generation and prior to simulation. 
The GEANT4 detector geometry
included a distribution of disabled pixel modules 
matching that in the experiment.
The MC events were used to derive correction factors from reconstructed
pixel tracks and two-point tracklets to the primary HIJING particles. 
Primary particles were defined to be either particles
originating directly from the \PbPb\ collision or particles
  resulting from secondary decays of
  HIJING produced particles with lifetimes $c\tau < 1$~cm. 

From the MC simulated events, correction factors accounting
for particle detection efficiency, misidentified tracks or tracklets
from unrelated clusters, 
and extra tracks or tracklets from secondary decays or from
interactions in the detector were calculated. The correction
  factors were evaluated in 20 intervals of detector
occupancy (\occ) parameterized using the number of reconstructed
clusters in the first pixel layer in the region $|\eta| < 1$.
Different corrections were applied to the pixel track and both
two-point tracklet measurements. For the pixel tracks, the 
efficiency, $\varepsilon_{\mathrm{pt}}$, for reconstructing 
tracks associated with charged primary particles was obtained from
\begin{equation}
\varepsilon_{\mathrm{pt}}(\occ, \eta) \equiv \frac{N^{\mathrm{match}}_{\mathrm{pr}}(\occ,
  \eta)}{N_{\mathrm{pr}}(\occ, \eta)},
\label{eq:pixtrkeffic}
\end{equation}
where $N_{\mathrm{pr}}$ represents the number of charged primary particles produced by
HIJING within a given $\eta$ interval, and $N^{\mathrm{match}}_{\mathrm{pr}}$ represents the
portion of those primary particles matched to reconstructed pixel
tracks. 
The contributions to the number of
reconstructed pixel tracks ($N_{\mathrm{pt}}$) from ``background''
sources were separately evaluated to produce a ``background''
fraction 
\begin{equation}
b_{\mathrm{pt}}(\occ, \eta) \equiv \frac{N^{\mathrm{backg}}_{\mathrm{pt}}(\occ, \eta)}{N_{\mathrm{pt}}(\occ,
\eta)}, 
\end{equation}
where $N^{\mathrm{backg}}_{\mathrm{pt}}$ represents the number of tracklets from
secondary interactions and decays, from particles initially produced outside the
kinematic acceptance of the measurement but scattering into it, 
and from combinations of clusters not associated
with any primary or secondary particle in the GEANT4 simulation.
This factor was combined with $\varepsilon_{\mathrm{pt}}(\occ, \eta)$ to
produce a correction factor
\begin{equation}
C_{\mathrm{pt}}(\occ, \eta) \equiv \frac{1}{\varepsilon_{\mathrm{pt}}(\occ, \eta)} \left(1 - b_{\mathrm{pt}}(\occ, \eta) \right).
\label{eq:pixtrktrkcorr}
\end{equation}
For the 0-10$\%$ centrality interval,
$\varepsilon_{\mathrm{pt}}$ is about 0.55 and $b_{\mathrm{pt}}$ 
is about 0.02 in the mid-rapidity region, giving 
a $C_{\mathrm{pt}}$ of about 1.8.

\begin{figure}
\begin{center}
\includegraphics[width=0.8\textwidth]{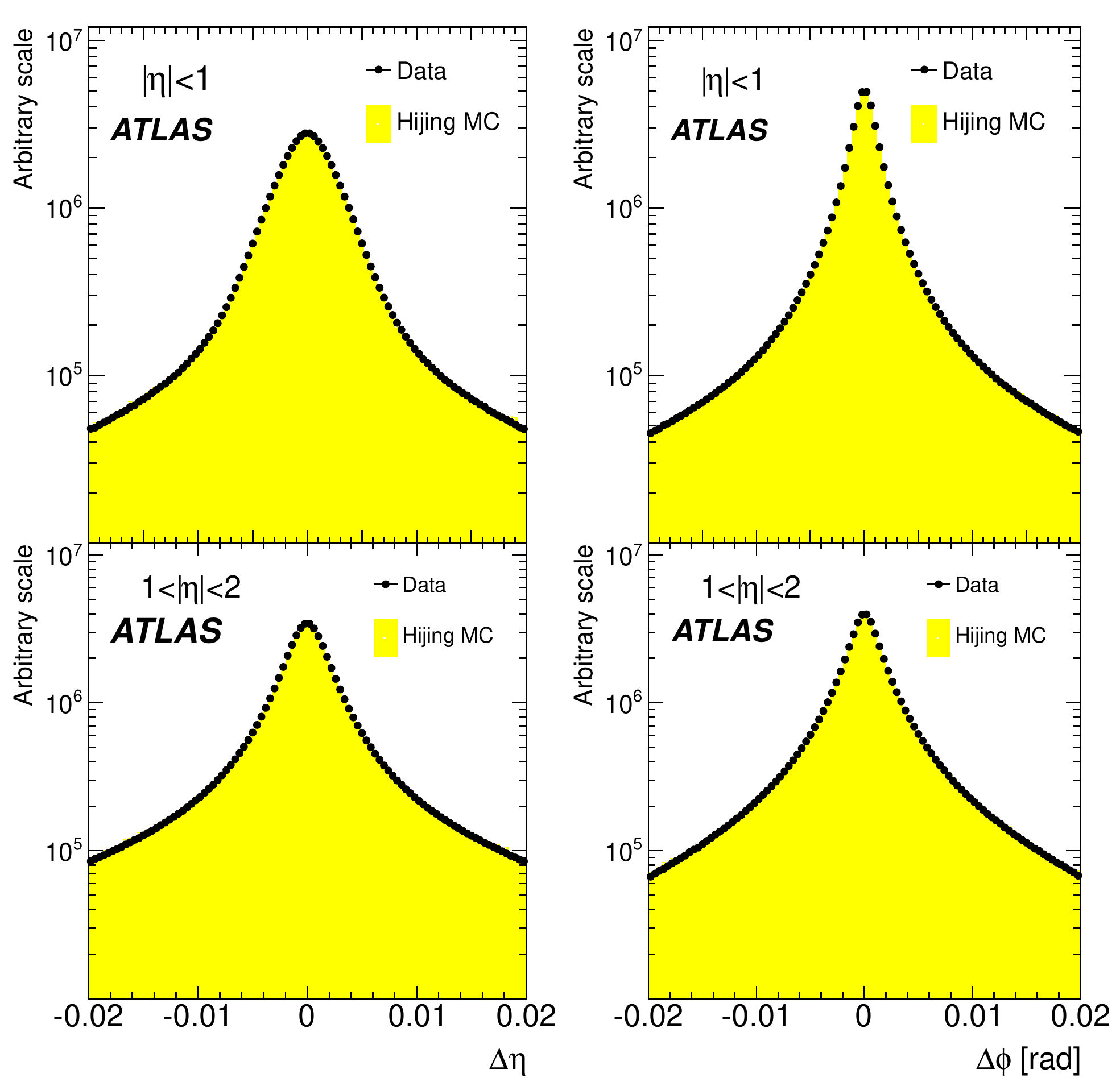}
\caption{Tracklet candidate $\Delta\eta$ (left)
and $\Delta\phi$ (right) distributions from data (histogram) and
reweighted MC (shaded region) for \PbPb\ collisions at $\sqrts=2.76$~TeV.  
The top panels correspond to $|\eta|<1$ 
and the bottom panels correspond to  $1<|\eta|<2$. 
Data and MC distributions are normalized to the same area. }
\label{fig:detadphi}
\end{center}
\end{figure}

For the two-point tracklet methods, a single multiplicative
correction factor was obtained from the MC simulations,
\begin{equation}
C_{\mathrm{2p}}(\occ, \eta) \equiv \frac{N_{\mathrm{pr}}(\occ, \eta)}{\Ntptr(\occ, \eta)},
\label{eq:twopttrkcorr}
\end{equation}
where $\Ntptr(\occ, \eta)$ represents reconstructed
tracklets. 
For the two-point tracklet Method~2, $N_{\mathrm{2p}}(\occ, \eta)$  was obtained
from the MC events via Eq.~\ref{eq:flipsub} using the same
flipping procedure as that applied in the data. For the 0-10$\%$ 
centrality interval, the correction factor is about 1.05 for Method~1
and 1.25 for Method~2 in the mid-rapidity region.

The \PbPb\ charged particle \pt\ spectrum measured at $\sqrts =
2.76$~TeV \cite{Aamodt:2010jd} differs from the spectrum generated by
HIJING at low and high \pt, with the generator exceeding the data
by 20\% at $\pT=500$ MeV, and underpredicting
  the charged particle yield by a factor of about two at $\pT=1.5$ GeV. 
Because the MC corrections
are applied to the data in matching \occ\ intervals, the mismatch in the
spectrum does not influence the corrections for misidentified tracks or 
occupancy-induced inefficiencies. However, if left uncorrected
the mismatch could distort the \pt-weighted single
track or tracklet efficiencies in the calculated correction
factors. To avoid this distortion a \pt-dependent
weight was applied to the generated particles  and to
tracklets or tracks that match generated 
particles in Eqs.~\ref{eq:pixtrkeffic}-\ref{eq:twopttrkcorr}. 
The \pt-dependent weights were obtained using an 
iterative procedure that,
in each analyzed centrality interval,
optimally matched 
the \pt\ spectrum of pixel tracks 
in \PbPb\ data with the solenoid magnet turned-on
to the reweighted spectrum produced
from a separate sample of HIJING+GEANT4 simulations
also performed with the solenoid turned-on.
Distributions of $\Delta
\eta$ and $\Delta \phi$ for candidate tracklets 
are shown in
Fig.~\ref{fig:detadphi} for two different pseudorapidity intervals,
$|\eta| < 1$ and $1 < |\eta| < 2$. The corresponding distributions
for the reweighted HIJING+GEANT4 events are also shown in
the figure and compare well with the data.
The maximum difference between data and MC is less than 5$\%$. 
It should be noted that the $\sigma_\eta(\eta)$ and $\sigma_\phi(\eta)$ mentioned above are evaluated
using the unreweighted MC, but they are applied consistently to data and reweighted MC
when calculating all $\eta$-dependent corrections.

\begin{figure}
\begin{center}
\includegraphics[width=1.0\textwidth]{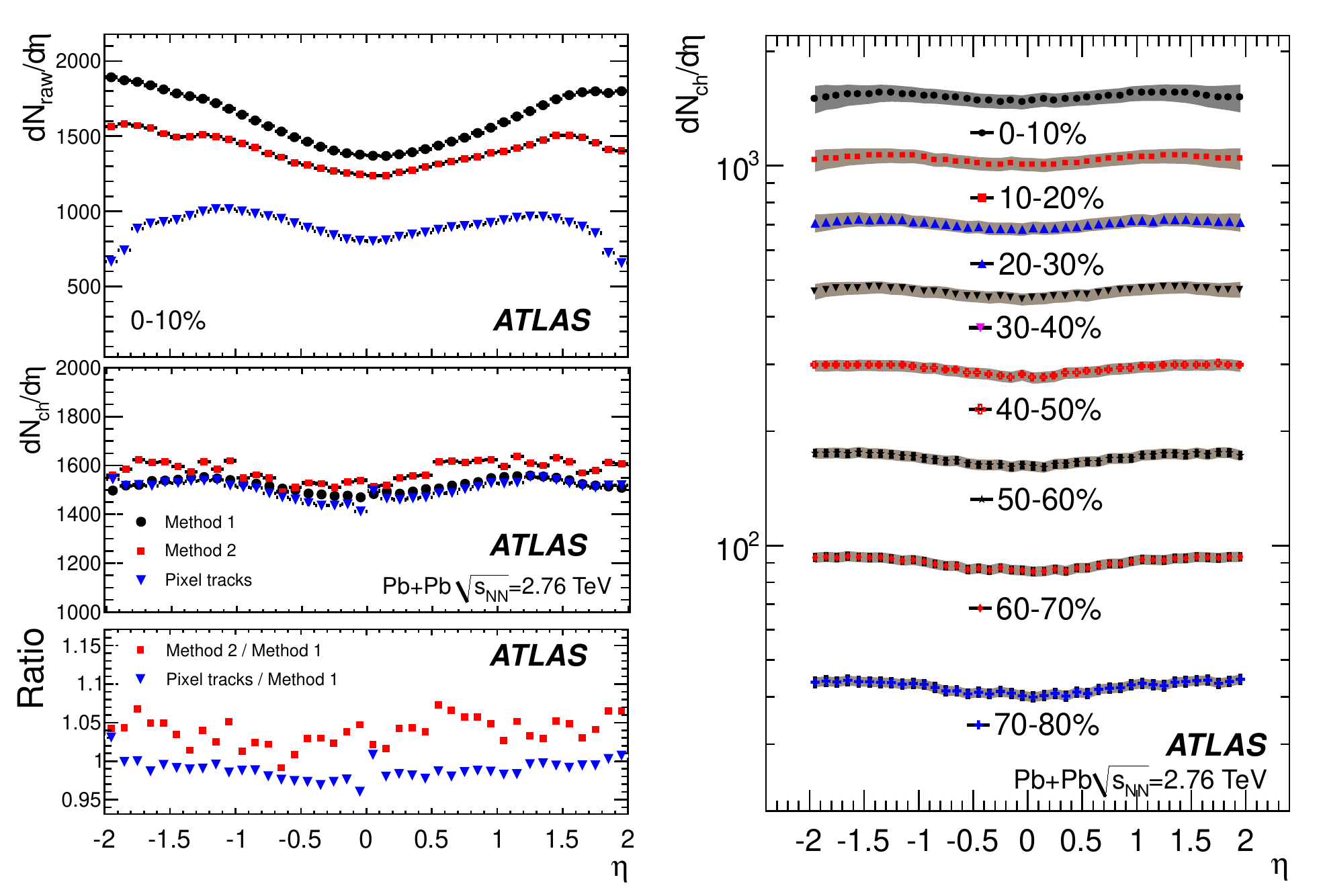}
\caption{{\bf Left:} Top: uncorrected track/tracklet
  $dN_{\mathrm{raw}}/d\eta$ distribution from tracklet Method~1 
(points), tracklet Method~2 (squares) and pixel tracking (blue
  triangles) for 0-10$\%$ centrality events. Middle: 
  corrected tracklet and track  \dndeta\ distributions.
  Bottom: ratio of \dndeta\ from
  the tracklet Method~2 (squares) and pixel tracking 
  (triangles) to tracklet Method~1. {\bf Right:} \dndeta\
  distributions from tracklet Method~1 for eight 10\% centrality intervals.
  The statistical errors are shown as bars and 
  the systematic errors are shown as shaded bands.
}
\label{fig:dndeta}
\end{center}
\end{figure}

Uncorrected pixel track and two-point tracklet pseudorapidity
distributions for 0-10\% centrality collisions are shown in the top left panel
of Fig.~\ref{fig:dndeta}. The corrections described above are
applied to obtain corrected, per-event primary charged particle
pseudorapidity distributions, averaged over the events in each centrality bin ($c$), according to
\begin{eqnarray}
\left. \frac{d\Nch}{d\eta}\right|_c = 
\frac{1}{\Nevt}\sum_{\mathrm{events},c} \frac{ \Delta \Nraw }{\Delta \eta} C(\occ, \eta),
\end{eqnarray}
where $\Delta \Nraw $ indicates either the number of reconstructed
pixel tracklets or two-point tracklets and $C(\occ, \eta)$ indicates
the $\eta$-dependent correction factors corresponding to the occupancy bin for each event. 
The corrected
\dndeta\
distributions for the 0-10\% centrality interval
are shown in the middle left panel of Fig.~\ref{fig:dndeta}. 
The bottom left panel of Fig.~\ref{fig:dndeta} shows the ratio of the
pixel tracking and two-point tracklet Method 2 results to the
two-point tracklet Method 1 results. In spite of
the factor of $\sim$2 differences between the raw yields for the three
reconstruction methods, the corrected pseudorapidity distributions
for central collisions agree within 5\%. The measurements
presented in the remainder of this paper were obtained from tracklet Method 1, 
which has the highest reconstruction efficiency, only a moderate
contribution of misidentified tracklets, and the smallest correction factors.
The resulting corrected \dndeta\ distributions are shown for 8 centrality intervals
in the right-hand panel of Fig.~\ref{fig:dndeta}. 

\section{Systematic uncertainties}

Various studies were performed to quantify the experimental 
uncertainties on the \dndeta\ measurement.
To address inaccuracies in the MC description of bad channels, 
disabled sensors, or other small instrumental problems, 
a comparison was made of unit-normalized $\eta$ and $\phi$
distributions of clusters in each of the first two pixel
layers between data and MC. The agreement between the $\eta$ and
$\phi$ distributions was found to be better than
0.05\% and 0.4\% in the first and second
  layers, respectively. Therefore, a combined
systematic uncertainty of 0.4\% is assigned to account for potential
MC inaccuracies.  
To evaluate the impact of inaccuracies in the description of the
detector material in the GEANT4 simulation, a separate set of HIJING+GEANT4
simulations was performed with an artificial 10\% increase in detector material and
a 15-20\% increase in material in various non-instrumented regions. 
The results obtained using correction factors from this ``extra
material'' sample agree with those obtained using the
default corrections to better than 2\%. 
Furthermore, the analysis was repeated using a different
\Rmc\ selection (see Eq.~\ref{eq:trackletcuts}), $\Rmc < 1.5$,
which should have a different sensitivity to multiple scattering,
secondaries, and occupancy effects. 
The corrections for the $\Rmc < 1.5$ selection differ from
those of the default analysis in central (0-10\%) collisions by 10\%
at $\eta = 0$ and 20\% at $\eta = 2$. However, the corrected
pseudorapidity distributions agree to 1\% in all centrality intervals.
 To address differences between 
the HIJING description of particle production in \PbPb\ collisions and
reality, the analysis was performed without the \pt\ spectrum re-weighting;
the results agree with those obtained using the re-weighting within
0.5\%. To address potential errors resulting from 
discrepancies in particle composition between data and MC,
the changes in
correction factors that would result from enhanced charged kaon
and proton production as observed at RHIC~\cite{Adler:2003cb} have
been evaluated. From the impact of the modified corrections on
the final result, a 1\% systematic uncertainty due to
incomplete knowledge of the hadron composition is assigned. 
To further test the sensitivity of the
  results to the use of the HIJING generator,
 a set of MC simulations using the HYDJET event
generator~\cite{HYDJET} was produced, and a separate set of correction
factors was obtained from this MC sample. 
HYDJET has a more complete description of soft particle production than HIJING, 
including a description of elliptic flow, and the version used here was tuned to have much lower multiplicities than found in HIJING.
In central collisions, the results 
obtained using the HYDJET-based corrections agree with
the HIJING-based results to better than 0.5\% at mid-rapidity, but
differ by as much as 7.5\% at $\eta = \pm 2$. A
centrality-dependent and $\eta$-dependent systematic error is assigned
to account for this difference. To address the inaccuracies from the
analysis procedure, a systematic uncertainty is assigned based on the
differences between the results obtained from the three reconstruction
methods described in this paper.  That uncertainty is
centrality-dependent and maximal for the 0-10\% centrality 
interval for which a 3.5\% uncertainty on
the overall scale of the pseudorapidity distribution is assigned based on the
comparison of the three results in the left, bottom panel of
Fig.~\ref{fig:dndeta}. 
The systematic uncertainties described above are summarized in
Table~\ref{tbl:systerr} for the most central (0-10\%) and 
the most peripheral (70-80\%) intervals. The total systematic uncertainties
are shown as shaded bands in the right panel of Fig.~\ref{fig:dndeta}.
\begin{table}
\begin{center}
\footnotesize
\begin{tabular}{|l|r|r|}\hline\hline
Source & Uncertainty (0-10\%) & (70-80\%)\\ \hline \hline
MC detector description & 0.4\% & 0.4\%\\ \hline
Extra material & 2\% & 2\% \\ \hline
$\Rmc$ cut & 1\% & 1\%\\ \hline
\pt\ re-weighting & 0.5\% & 0.5\%\\ \hline
Hadron composition & 1\% & 1\%\\ \hline
Enhanced $K_s$, $\Lambda$ & 1\% & 1\%\\ \hline
HYDJET  & 0.5-7.5\% vs. $\eta$ & 0\%\\ \hline
Analysis Method  & 3.5\% & 1\%\\ \hline \hline
Combined ($\eta = 0$)& 4\% & 3\%\\ \hline 
Combined ($\eta = 2$)& 8.5\% & 3\%\\ \hline \hline
\end{tabular}
\caption{Summary of the various sources of systematic uncertainties 
  and their estimated impact
  on the \dndeta\ measurement in central (0-10\%) and peripheral (70-80\%)
  \PbPb\ collisions. Only the uncertainty due to the choice of the 
event generator is 
  $\eta$-dependent.
}
\label{tbl:systerr}
\end{center}
\end{table}

\section{Results}

The measured charged particle \dndeta\ shown in Fig.~\ref{fig:dndeta}, increases rapidly with
collision centrality for all $\eta$. It is conventional to characterize 
particle production in nucleus-nucleus collisions by the mid-rapidity
\dndeta, \dndetamid, which here is defined to be  
\dndeta\ averaged over $|\eta|<0.5$. The analysis presented in this paper yields
\dndetamid\ values in central collisions of $1479 \pm 10 {\rm
  ({\rm stat.})} \pm 63 ({\rm syst.})$,  $1598 \pm 11 {\rm
  ({\rm stat.})} \pm 68 ({\rm syst.})$, and $1738 \pm 12 {\rm (stat.)} \pm 75
({\rm syst.})$  for the 0-10\%, 0-6\%, and 0-2\% centrality intervals, 
respectively. 
Table~\ref{table:npart} provides results of the 
\dndetamid\ measurements for all centrality bins. 

\begin{figure}
\begin{center}
\includegraphics[width=0.95\textwidth]{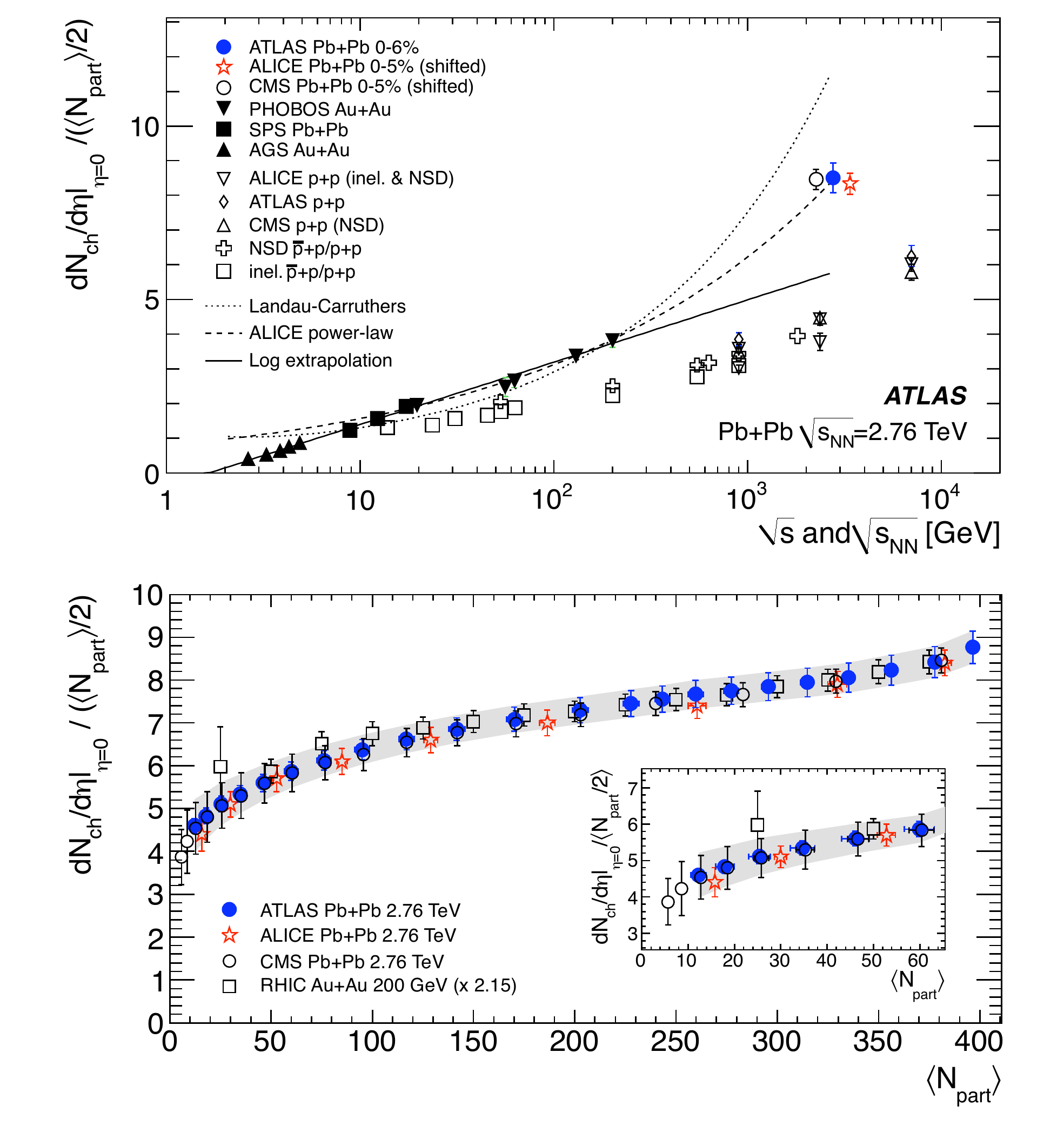}
\caption{ 
{\bf Top:} 
\sqrts\ dependence of the charged particle \dndeta\ per 
colliding nucleon pair $\dndetamid/(\Npart/2)$ from a variety of measurements in
\pp\  and \mbox{$\bar{\mathrm{p}}$+p} (inelastic and non-single diffractive results from \cite{Alver:2010ck} and
references therein,
as well as ~\cite{:2010ir}-\cite{:2009dt}) 
and central \nucnuc\ collisions,
including the ATLAS 0-6\% centrality measurement reported
here for $|\eta|<0.5$ and the previous 0-5\% centrality ALICE~\cite{Collaboration:2010cz} 
and CMS~\cite{CMS:2011mult} measurements (points shifted 
horizontally for clarity). 
The curves show different expectations for
the \sqrts\ dependence in \nucnuc\ collisions:
results of a Landau hydrodynamics calculation~\cite{Carruthers:1973ws} (dotted line)
, an $s^{0.15}$ extrapolation of RHIC and SPS data proposed by
ALICE \cite{Collaboration:2010cz} (dashed line), a logarithmic
extrapolation of RHIC and SPS data from \cite{Busza:2007ke} (solid line).
{\bf Bottom:} $\dndetamid/(\Npart/2)$ 
  vs \Npart\ for 2\% centrality intervals over 0-20\% and 5\% centrality
  intervals over 20-80\%. Error bars represent combined statistical and
  systematic uncertainties on the \dndetamid\ measurements, whereas the shaded band
  indicates the total systematic uncertainty including
  \Npart\ uncertainties. The RHIC measurements (see text) have been multiplied by
  2.15 to allow comparison with the $\sqrts = 2.76$~\TeV\ results.
  The inset shows the $\langle N_{\mathrm{part}} \rangle<60$ region in more detail. 
  }
\label{fig:physplot}
\end{center}
\end{figure}

The top panel of Fig.~\ref{fig:physplot} compares the ATLAS measurement to the previously
reported ALICE~\cite{Collaboration:2010cz} and CMS~\cite{CMS:2011mult} results 
for $|\eta|<0.5$ for
the 0-5\% centrality interval in terms of
\dndetamid\ per colliding nucleon pair, $\dndetamid/(\Npart/2)$, 
and to other \nucnuc\ measurements at different
\sqrts\ (see \cite{Alver:2010ck}, which includes data from Refs.\cite{468861}-\cite{644319}).
The ALICE and CMS
0-5\% centrality measurements agree with the result reported here
for the 0-6\% centrality interval, $8.5 \pm 0.1 {\rm (stat.)} \pm 0.4 (\rm syst.)$, within the quoted errors. 
The LHC results show that the multiplicity
in central \nucnuc\ collisions
rises rapidly with \sqrts\ above the RHIC top energy of \sqrts=200 GeV. The three
curves shown in Fig.~\ref{fig:physplot} indicate possible variations of
$\dndetamid/(\Npart/2)$ with \sqrts. The dotted curve describes a
\sqrts\ dependence expected from Landau hydrodynamics
\cite{Carruthers:1973ws}.
It is clearly inconsistent with the data.
The dot-dashed curve represents a logarithmic extrapolation of RHIC and
SPS data \cite{Busza:2007ke} that is also excluded by the
measurement presented in this paper and by the ALICE and CMS measurements.
The dashed curve shows an $s^{0.15}$ dependence suggested by
ALICE \cite{Collaboration:2010cz} that is consistent
with the ATLAS measurement. Also shown in the top panel in Fig.~\ref{fig:physplot} are
results from \pp\ and \mbox{$\bar{\mathrm{p}}$+p} measurements at different $\sqrt{s}$ (\cite{Alver:2010ck} and references therein, 
as well as~\cite{:2010ir}-\cite{:2009dt}). The excess of $\dndetamid/(\Npart/2)$ in
\nucnuc\ collisions over \pp\ collisions observed at RHIC
 persists and is
proportionately larger at the higher \sqrts\ values of the LHC. 

\begin{table}[!tbp]
  \centering
\footnotesize
  \begin{tabular}{|c | |c| c| c|}
    \hline \hline
    Centrality & $<N_{part}>$ & \dndetamid & $\dndetamid/\Npart/2$ \\
    \hline
    0-2$\%$   & 396 $\pm$ 2 & 1738 $\pm$ 76 & 8.8 $\pm$ 0.4 \\
    2-4$\%$   & 378 $\pm$ 2 & 1591 $\pm$ 67 & 8.4 $\pm$ 0.4 \\
    4-6$\%$   & 356 $\pm$ 3 & 1467 $\pm$ 63 & 8.2 $\pm$ 0.4 \\
    6-8$\%$   & 335 $\pm$ 3 & 1350 $\pm$ 57 & 8.1 $\pm$ 0.4 \\
    8-10$\%$  & 315 $\pm$ 3 & 1250 $\pm$ 53 & 8.0 $\pm$ 0.3 \\
    \hline
    10-12$\%$ & 296 $\pm$ 3 & 1159 $\pm$ 48 & 7.8 $\pm$ 0.3 \\
    12-14$\%$ & 277 $\pm$ 4 & 1074 $\pm$ 44 & 7.8 $\pm$ 0.3 \\
    14-16$\%$ & 260 $\pm$ 4 & 996 $\pm$ 41  & 7.7 $\pm$ 0.3 \\
    16-18$\%$ & 243 $\pm$ 4 & 918 $\pm$ 37  & 7.6 $\pm$ 0.3 \\
    18-20$\%$ & 228 $\pm$ 4 & 849 $\pm$ 34  & 7.5 $\pm$ 0.3 \\
    \hline
    20-25$\%$ & 203 $\pm$ 4 & 739 $\pm$ 29  & 7.3 $\pm$ 0.3 \\ 
    25-30$\%$ & 170 $\pm$ 4 & 603 $\pm$ 24  & 7.1 $\pm$ 0.3 \\ 
    \hline
    30-35$\%$ & 142 $\pm$ 4 & 486 $\pm$ 19  & 6.9 $\pm$ 0.3 \\ 
    35-40$\%$ & 117 $\pm$ 4 & 387 $\pm$ 15  & 6.6 $\pm$ 0.3 \\ 
    \hline
    40-45$\%$ & 95.0 $\pm$ 3.7  & 303 $\pm$ 11  & 6.4 $\pm$ 0.3 \\ 
    45-50$\%$ & 76.1 $\pm$ 3.5  & 233 $\pm$ 9   & 6.1 $\pm$ 0.4 \\ 
    \hline
    50-55$\%$ & 59.9 $\pm$ 3.3  & 176 $\pm$ 6   & 5.9 $\pm$ 0.4 \\ 
    55-60$\%$ & 46.1 $\pm$ 3.0  & 129 $\pm$ 5   & 5.7 $\pm$ 0.4 \\ 
    \hline
    60-65$\%$ & 34.7 $\pm$ 2.7  & 93 $\pm$ 3    & 5.3 $\pm$ 0.5 \\ 
    65-70$\%$ & 25.4 $\pm$ 2.3  & 65 $\pm$ 2    & 5.1 $\pm$ 0.5 \\ 
    \hline
    70-75$\%$ & 18.0 $\pm$ 2.0  & 43 $\pm$ 2    & 4.8 $\pm$ 0.6 \\ 
    75-80$\%$ & 12.3 $\pm$ 1.6  & 28 $\pm$ 1    & 4.6 $\pm$ 0.6 \\ 
    \hline
    \hline
  \end{tabular}
  \caption{Tabulation of measurements of \dndetamid\ evaluated over $|\eta|
    < 0.5$ and
    $\dndetamid/(\Npart/2)$ for the full set of
    centrality bins considered in the analysis and shown in
    Fig.~\ref{fig:physplot}. The uncertainties on \dndetamid\ include
    statistical and systematic errors on the multiplicity
    measurement. The errors reported for $\dndetamid/(\Npart/2)$ also
    include systematic uncertainties on the centrality selection and
    \Npart\ determination.
  } 
  \label{table:npart}
\end{table}

The bottom panel of Fig.~\ref{fig:physplot} shows 
$\dndetamid/(\Npart/2)$ as a function
of \Npart\ for 2\% centrality intervals over 0-20\%, and 5\% centrality
intervals over 20-80\%. The values are also reported in
Table~\ref{table:npart}. A moderate variation of $\dndeta|_{\eta = 0}/(\Npart/2)$ with \Npart\ is
observed, from a value of $4.6 \pm 0.1 ({\rm stat.}) \pm 0.6 ({\rm syst.})$ at
$\Npart = 12.3$ (centrality 75-80\%) to $8.8 \pm 0.1 ({\rm stat.}) \pm 0.4 ({\rm syst.})$ at
$\Npart = 396$ (centrality 0-2\%). The increase of
$\dndetamid/(\Npart/2)$ with $\Npart$ is monotonic 
up to the most central interval (0-2\%).
This demonstrates that, even for the most central collisions, variations in
centrality -- as characterized by 
transverse energy depositions well outside the
acceptance used for the multiplicity measurement -- yield significant changes 
in the measured final state multiplicity. 

The bottom panel of Fig.~\ref{fig:physplot} also shows 
ALICE and CMS measurements of \dndetamid\ as a function of \Npart\  
that agree with the results presented here for all centrality intervals. 
Also shown are results from \AuAu\ collisions at $\sqrts = 200$~GeV
obtained from an average of measurements
from the four RHIC collaborations~\cite{Adler:2004zn}-\cite{Adams:2004cb}. 
Similar to the approach used in Ref.~\cite{Collaboration:2010cz}, the
200~GeV \AuAu\ results have been scaled by a factor of $2.15$ to allow
comparison with the $\sqrts = 2.76$~\TeV\ data. This
factor was obtained by matching the most central 200~GeV
\AuAu\ \dndeta\ measurement at $\eta = 0$ to 
the \dndeta\ measurement from this paper at $\eta = 0$ in the 2-4\% centrality
interval, the interval that has the closest value of \Npart\ to the
most central 200~GeV measurement.
After re-scaling, the trend of the 200~GeV data is 
in good agreement with the 2.76~TeV measurements for all
reported centrality intervals. Similar
observations have been made previously in comparisons of top energy
RHIC data to much lower energies~\cite{Alver:2010ck}. Therefore, this scaling behavior appears 
to be a robust feature of particle production in heavy ion collisions.  

\begin{figure}
  \begin{center}
    \includegraphics[width=1.0\textwidth]{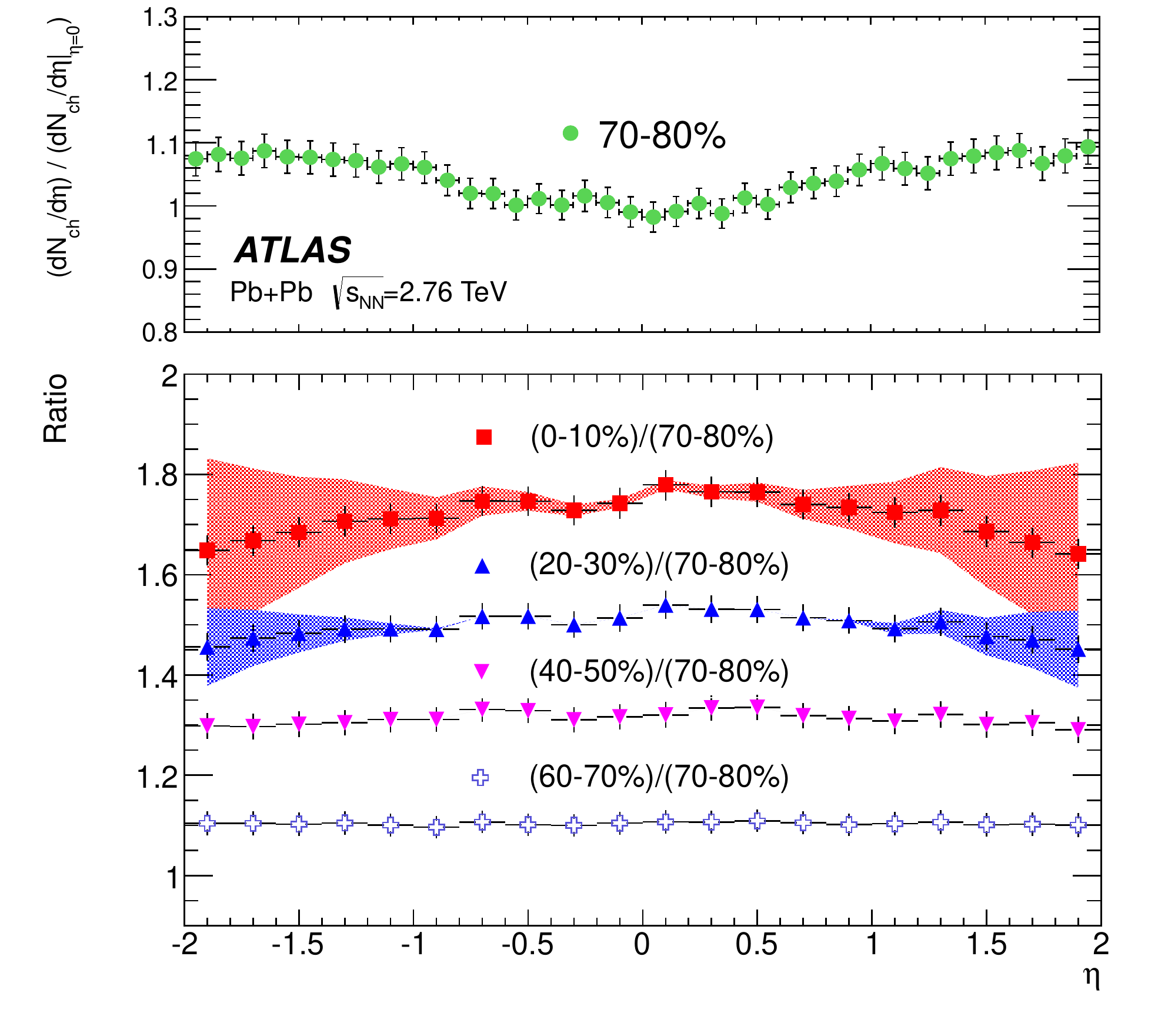} 
    \caption{{\bf Top:} \dndeta\
      distributions from tracklet Method~1, scaled by \dndetamid, as a function
      of the pseudorapidity for the 70-80\% centrality interval.
      The statistical errors are shown as error bars.
      {\bf Bottom:} Ratio of $\dndeta/(\Npart/2)$ measured 
      in different centrality intervals: 0-10\% (squares), 
      20-30\% (triangles), 40-50\% (inverted triangles) and 60-70\% (crosses) 
      to that measured in peripheral collisions (70-80\%).
      Statistical uncertainties are shown as bars while
      $\eta$-dependent systematic uncertainties are shown as shaded bands. }
    \label{fig:ratio}
  \end{center}
\end{figure}

To evaluate the shapes of the measured charged particle \dndeta\ distributions
Fig.~\ref{fig:ratio} (top) shows the \dndeta\ distribution divided by
\dndetamid\ for the 70-80\% centrality interval. For this centrality interval, 
the \dndeta\ increases by 7\% $\pm$ 1\% from 
$\eta = 0$  to $|\eta| > 1$. The bottom panel
shows ratios of \dndetaperpair\ for several other 10\%
centrality intervals to the same quantity in the 70-80\% interval. 
No significant variation of the shape of \dndeta\ with centrality is
observed within the 
systematic uncertainties. 

\section{Conclusions}

This paper presents results on the measurement of charged
particle pseudorapidity distributions 
over $|\eta| <2$ as a function of collision centrality
in a sample of $\sqrts = 2.76$~TeV lead-lead collisions
recorded with the ATLAS detector at the LHC.
Three different analysis methods are used, based on the pixel detector and
using events with the solenoid magnet turned off in order to measure
particles with transverse momenta as low as 30 MeV.
The charged particle mid-rapidity \dndeta, normalized by $\Npart/2$, is found to increase
significantly with beam energy by about a factor of two relative to 
earlier RHIC data, and is substantially larger than $\pp$ data at the same
energy.
The relative centrality dependence of
$\dndeta|_{\eta = 0}/(\Npart/2)$ 
agrees well with that observed at RHIC.
These results agree well with previous
mid-rapidity measurements from ALICE and CMS. 
Furthermore, the peripheral (70-80\%) $\dndeta$ distribution shows a significant
rise with increasing $|\eta|$ away from $\eta =
0$. No variation of the shape of the \dndeta\ distribution with
centrality outside the reported systematic uncertainties is observed.

\section*{Acknowledgments}

\input{Acknowledgement-29Nov10-collisions.tex}

\include{Biblio}
\clearpage

\label{app:ATLASColl}
 \begin{center}
 \input{atlas_authlist}
 \end{center}

\clearpage

\end{document}

%% file: Acknowledgement-29Nov10-collisions.tex
We  thank CERN for the efficient commissioning and operation of the LHC during this initial heavy ion data taking period as well as the support staff from our institutions without whom ATLAS could not be operated efficiently.

We acknowledge the support of ANPCyT, Argentina; YerPhI, Armenia; ARC, Australia; BMWF, Austria; ANAS, Azerbaijan; SSTC, Belarus; CNPq and FAPESP, Brazil; NSERC, NRC and CFI, Canada; CERN; CONICYT, Chile; CAS, MOST and NSFC, China; COLCIENCIAS, Colombia; MSMT CR, MPO CR and VSC CR, Czech Republic; DNRF, DNSRC and Lundbeck Foundation, Denmark; ARTEMIS, European Union; IN2P3-CNRS, CEA-DSM/IRFU, France; GNAS, Georgia; BMBF, DFG, HGF, MPG and AvH Foundation, Germany; GSRT, Greece; ISF, MINERVA, GIF, DIP and Benoziyo Center, Israel; INFN, Italy; MEXT and JSPS, Japan; CNRST, Morocco; FOM and NWO, Netherlands; RCN, Norway;  MNiSW, Poland; GRICES and FCT, Portugal;  MERYS (MECTS), Romania;  MES of Russia and ROSATOM, Russian Federation; JINR; MSTD, Serbia; MSSR, Slovakia; ARRS and MVZT, Slovenia; DST/NRF, South Africa; MICINN, Spain; SRC and Wallenberg Foundation, Sweden; SER,  SNSF and Cantons of Bern and Geneva, Switzerland;  NSC, Taiwan; TAEK, Turkey; STFC, the Royal Society and Leverhulme Trust, United Kingdom; DOE and NSF, United States of America.  

The crucial computing support from all WLCG partners is acknowledged gratefully, in particular from CERN and the ATLAS Tier-1 facilities at TRIUMF (Canada), NDGF (Denmark, Norway, Sweden), CC-IN2P3 (France), KIT/GridKA (Germany), INFN-CNAF (Italy), NL-T1 (Netherlands), PIC (Spain), ASGC (Taiwan), RAL (UK) and BNL (USA) and in the Tier-2 facilities worldwide.

%% file: atlas_authlist.tex
\begin{flushleft}
{\Large The ATLAS Collaboration}

\bigskip

G.~Aad$^{\rm 48}$,
B.~Abbott$^{\rm 111}$,
J.~Abdallah$^{\rm 11}$,
A.A.~Abdelalim$^{\rm 49}$,
A.~Abdesselam$^{\rm 118}$,
O.~Abdinov$^{\rm 10}$,
B.~Abi$^{\rm 112}$,
M.~Abolins$^{\rm 88}$,
H.~Abramowicz$^{\rm 153}$,
H.~Abreu$^{\rm 115}$
% ...
\bigskip

% ...
$^{10}$ Institute of Physics, Azerbaijan Academy of Sciences, Baku, Azerbaijan\\
$^{11}$ Institut de F\'isica d'Altes Energies and Departament de F\'isica de la Universitat Aut\`onoma  de Barcelona and ICREA, Barcelona, Spain\\
% ...
$^{48}$ Fakult\"{a}t f\"{u}r Mathematik und Physik, Albert-Ludwigs-Universit\"{a}t, Freiburg i.Br., Germany\\
$^{49}$ Section de Physique, Universit\'e de Gen\`eve, Geneva, Switzerland\\
% ...
$^{88}$ Department of Physics and Astronomy, Michigan State University, East Lansing MI, United States of America\\
% ...
$^{111}$ Homer L. Dodge Department of Physics and Astronomy, University of Oklahoma, Norman OK, United States of America\\
$^{112}$ Department of Physics, Oklahoma State University, Stillwater OK, United States of America\\
% ...
$^{115}$ LAL, Univ. Paris-Sud and CNRS/IN2P3, Orsay, France\\
% ...
$^{118}$ Department of Physics, Oxford University, Oxford, United Kingdom\\
% ...
$^{153}$ Raymond and Beverly Sackler School of Physics and Astronomy, Tel Aviv University, Tel Aviv, Israel\\
% ...
\end{flushleft}


%% file: atlas_dndetaHI_plb.bbl
\begin{thebibliography}{00}


\bibitem{LHCHI}
  P. Giubellino,
 arXiv:0809.1062 [nucl-ex].

\bibitem{Alver:2010ck}
  PHOBOS Collaboration, 
  Phys.~Rev.~ {\bf C83} (2011) 024913.




\bibitem{Wang:1991hta}
 X.~-N.~Wang and M.~Gyulassy,
 Phys.\ Rev.\  {\bf D44 } (1991) 3501, HIJING version 1.38b with no additional tuning.

\bibitem{Deng:2010mv}
  W.~-T.~Deng, X.~-N.~Wang and R.~Xu,
  Phys.\ Rev.\ C {\bf 83} (2011) 014915.

\bibitem{ALbacete:2010ad}
  J.~L.~Albacete and A.~Dumitru,
  arXiv:1011.5161 [hep-ph].

\bibitem{Levin:2010dw}
  E.~Levin and A.~H.~Rezaeian,
  Phys.\ Rev.\  {\bf D82 } (2010) 014022.

\bibitem{Carruthers:1973ws}
  P.~Carruthers and M.~Doung-van,
  Phys.\ Rev.\  {\bf D8 } (1973) 859.

\bibitem{Collaboration:2010cz}
  ALICE Collaboration, 
  Phys.~Rev.~Lett. {\bf 106} (2011) 032301.

\bibitem{CMS:2011mult}
  CMS Collaboration, 
  JHEP {\bf 1108} (2011) 141.

\bibitem{AtlasDetector}
  ATLAS Collaboration, 
 JINST {\bf 3} (2008) S08003.



\bibitem{atlasdetector}
  ATLAS Collaboration, 
  JINST {\bf 3} (2008) S08003.

\bibitem{pixel}
  ATLAS Collaboration, 
  JINST {\bf 3} (2008) P07007.

\bibitem{Djuvsland:2010qs}
  O.~Djuvsland and J.~Nystrand,
  Phys.\ Rev.\  {\bf C83} (2011) 041901.

\bibitem{Miller:2007ri}
 M.~L.~Miller {\it et al.},
 Ann.\ Rev.\ Nucl.\ Part.\ Sci.\  {\bf 57 } (2007) 205.


\bibitem{Alver:2008aq}
 B.~Alver {\it et al.},
 arXiv:0805.4411 [nucl-ex].


\bibitem{:2010px}
  ATLAS Collaboration, Phys.\ Lett.\  {\bf B697 } (2011) 294.

\bibitem{Nakamura:2010zzi}
  K.~Nakamura {\it et al.} (Particle Data Group Collaboration),  J.\ Phys.\ {\bf G37 } (2010) 075021.

\bibitem{Newtracking1} T. Cornelissen {\it et al.}, 
  J.\ Phys.\ Conf.\ Ser.\  {\bf 119 } (2008)  032014.


\bibitem{Agostinelli:2002hh}
 GEANT4 Collaboration, S.~Agostinelli {\it et al.},
 Nucl.\ Instrum.\ Meth.\ {\bf A506} (2003) 250.

\bibitem{ATLASSim}
 ATLAS Collaboration, 
 Eur. Phys. J.  {\bf C70} (2010) 823.
 
\bibitem{Aamodt:2010jd}
  ALICE Collaboration, 
  Phys.\ Lett.\  {\bf B696} (2011)  30.

\bibitem{Adler:2003cb}
  PHENIX Collaboration, 
Phys.~Rev. {\bf C64} (2004)  034909.

\bibitem{HYDJET}
  I.P. Lokhtin and A.M. Snigirev, Eur. \ Phys. \ J. \ {\bf C46} (2006) 211.

\bibitem{468861} 
  E802 Collaboration, 
  Phys.\ Rev.\ {\bf C57} (1998) 466.

\bibitem{nucl-ex/0008010} 
  E866 and E917 Collaborations, 
  Phys.\ Lett.\ {\bf B490} (2000) 53.

\bibitem{arXiv:nucl-ex/0205002}
  NA49 Collaboration, 
  Phys.\ Rev.\ {\bf C66} (2002) 054902.

\bibitem{602292}
  PHOBOS Collaboration, 
  Phys.\ Rev.\ {\bf C66} (2002) 054901.

\bibitem{nucl-ex/0306033}
  E895 Collaboration, 
  Phys.\ Rev.\ {\bf C68} (2003) 054905.

\bibitem{644319}
  NA49 Collaboration, 
  Phys.\ Rev.\ C\ {\bf 69} (2004) 024902.

\bibitem{Busza:2007ke}
  W.~Busza,  J.\ Phys. {\bf G35 } (2008)  044040,\\
  PHOBOS Collaboration, 
  Phys.\ Rev.\  {\bf C74} (2006) 021901.

\bibitem{:2010ir}
  ATLAS Collaboration, 
  New J.\ Phys.\  {\bf 13} (2011) 053033.

\bibitem{Khachatryan:2010xs}
  CMS Collaboration, 
  JHEP {\bf 1002}  (2010) 041.

\bibitem{Khachatryan:2010us}
  CMS Collaboration, 
  Phys.\ Rev.\ Lett.\  {\bf 105} (2010) 022002.

\bibitem{Aamodt:2010pp}
  ALICE Collaboration, 
  Eur.\ Phys.\ J.\  {\bf C68}  (2010) 345.

\bibitem{:2009dt}
  ALICE Collaboration, 
  Eur.\ Phys.\ J.\   {\bf C68} (2010) 89.

\bibitem{Adler:2004zn}
  PHENIX Collaboration, 
  Phys.\ Rev.\   {\bf C71}  (2005) 034908 [Erratum-ibid.\  {\bf C71} (2005)  049901].

\bibitem{Bearden:2001xw}
  BRAHMS Collaboration, 
  Phys.\ Lett.\ {\bf B523} (2001) 227.

\bibitem{Bearden:2001qq}
  BRAHMS Collaboration, 
  Phys.\ Rev.\ Lett.\  {\bf 88} (2002) 202301.

\bibitem{Back:2002uc}
  PHOBOS Collaboration, 
  Phys.\ Rev.\  {\bf C65} (2002) 061901.

\bibitem{Adams:2004cb} 
  STAR Collaboration, 
  Phys.\ Rev.\ {\bf C70} (2004) 054907.






\end{thebibliography}
